%% file: main.tex
\documentclass[11pt]{article}
\usepackage[margin=1in]{geometry}
\usepackage[numbers,sort&compress]{natbib}
\usepackage{ccfinance-manuscript}
\usepackage[hidelinks]{hyperref}
\input{metadata}
\title{\papertitle}
\author{\paperauthors}
\date{\paperdate}
\begin{document}
\maketitle
\begin{abstract}
  \input{sections/00_abstract}
\end{abstract}
\input{body}

\bibliographystyle{plainnat}
\ifmanuscriptcites
\bibliography{refs}
\fi
\appendix
\renewcommand{\theHtable}{appendix.\arabic{table}}
\input{appendix}

\end{document}

%% file: metadata.tex
\newcommand{\papertitle}{CCFinance: Reconstructing Financial News
Samples from Web Archives}
\newcommand{\paperauthors}{Marcus Gawronsky}
\newcommand{\paperdate}{}

%% file: sections/00_abstract.tex
Proprietary news archives assign stories to firms, so reproducing a
financial study often requires access to both the articles and the
vendor's assignments. CCFinance lets researchers reconstruct firm coverage
under different selection rules by recording which version of a story
was captured, which issuers it names and how often each appears. From
14.8 million
English-language records in Fine News, derived from public Common Crawl
News, we identify 5.22 million distinct, consistently dated articles
with issuer links. Exact reconstruction requires the restricted snapshot
and derived artifacts. We compare GLiNER and Stanza on 250 model-labeled
article leads and assess issuer links and story focus against blinded
model judgments. Estimated precision among 475 accepted links is 96.8\%.
Among held-out articles with a detected company, requiring three
mentions raises subject precision from 31.9\% to 66.1\%. Across the
2017--2025 panel, the same rule reduces the number of issuer-years with
at least 50 qualifying stories from 13,918 to 8,641. Three
reimplementations examine
volatility tracking, bank networks and overnight returns. In the network
study, bank pairs appearing together rose from 177 to 242 across
publishers between January and April 2020, but fell from 50 to 35
within Reuters.

%% file: body.tex
\input{sections/01_introduction}
\input{sections/02_literature_review}
\input{sections/03_methodology}
\input{sections/04_empirical_design_and_data}
\input{sections/05_results}
\input{sections/06_conclusion}

%% file: sections/01_introduction.tex
\section{Introduction}
\label{sec:introduction}

Because much of the financial news used in empirical research comes
from proprietary archives, researchers seeking to reproduce a published
estimate often need access both to the original stories and to the
vendor annotations that associate those stories with firms.
\citet[Section~2]{heston2017}, for example, use timestamps and story IDs
to match proprietary NewsScope sentiment scores and firm tags to full
text from the Reuters News Archive. Even with a copy of the same story,
two researchers can construct different firm samples if one counts every
company named in an earnings report and the other counts only the firm
whose earnings it reports. Earlier collections available to researchers
have often covered fewer publishers or shorter periods, while the fields
needed to reconstruct a sample may be distributed across products.
The public SEntFiN release, for example, provides entity sentiment
annotations for 10,753 headlines \citep{sinha2022sentfin}, whereas a
study of what investors could observe may also require the article
body, its publication and capture times, and the identity of the listed
issuer at that date.

Researchers working outside a commercial subscription have generally
obtained news in one of two ways, either through a bounded release from
a publisher or vendor, or by collecting articles directly. NIST
distributes the Reuters Corpus Volume~1 and the Thomson Reuters
Text Research Collection (TRC2) under research agreements. The former
covers a year of English-language newswire and the latter the period
from January 2008 to February 2009 \citep{reuterscorpora}. Collecting
articles from publishers
or aggregators, as \citet{dong2024} do using Nasdaq's per-symbol pages,
gives researchers control over the material they retain, although the
pages available at the time of collection and the terms governing their
use determine how much of the sample others can recover.
\citet{remyding2015}, for example, document the withdrawal of their
original Reuters and Bloomberg downloads on copyright grounds.

Because direct collection revisits live sites, researchers must respect
publisher terms and crawl preferences while limiting the load they place
on servers. Controls that protect a site from abusive traffic can
challenge a research crawler even when the site admits verified search
crawlers \citep{cloudflare2026}, whose referrals can bring publishers
readers and advertising revenue \citep{cma2020publishers}. A shared
archive allows several teams to examine the same response after a single
fetch. This matters for replication because two visits to the same URL
can return different texts. If a publisher adds a second company's
response to a story after publication, researchers collecting on
different days can disagree even on which firms the story names. When
both researchers can recover the same capture, they can instead examine
how their respective rules treat the same passage. Common Crawl, for
example, observes publisher crawl directives
and slows its requests after rate limits or server errors
\citep{commoncrawlfaq}.

The scale at which such an archive can be reused is evident in language
model pretraining, where filtered Common Crawl data accounted for 60\%
of the GPT-3 training mixture by sampling weight
\citep[Table~2.2]{brown2020}. The nonprofit Common Crawl Foundation
makes both its URL index and the underlying web responses available to
researchers, policymakers and the public
\citep{commoncrawlabout,commoncrawloverview}. While commercial search
services index pages to retrieve and rank results \citep{google2026search},
the public archive also lets a researcher inspect a particular version
of a page after the live site has changed. In 2016 the Foundation
introduced CC-News, a worldwide crawl of news sites whose daily Web
ARChive (WARC) files record the fetched responses and their capture
times \citep{nagel2016}. Access to those copies remains subject to the
source terms and the rights in their content \citep{commoncrawlterms}.

Before an archived response can serve as article text, a researcher
has to identify its body and decide which copies to retain.
RefinedWeb illustrates the role of filtering and deduplication in
preparing web data for language model pretraining \citep{penedo2023}.
For news, Infini-News extracts article bodies and structured metadata
from CC-News and indexes them with their capture provenance
\citep{lazzaroni2026}, while Fine News applies language and quality
filters to those records and removes duplicate texts \citep{solovev2026}.
These transformations allow a researcher who starts with a cleaned
body to trace it back through the extraction to an archived response.
For a language model the body can already serve as a training example.
For a financial panel a further decision remains, because prices and
returns are observed for particular securities on particular dates.

Even after this processing, a single article can name several companies,
report an event that concerns only one, or express different sentiment
toward each \citep{shah2024,sinha2022sentfin}. Named-entity recognition
locates the mentions, and issuer linking assigns them dated identities
that a researcher can join to financial records \citep{schwenkler2020}.
Article-form tags distinguish, for example, a news report from a press
release, while event labels identify the kind of development a passage
describes. Its implications for investors can depend on whose shares
they hold, which gives issuer-specific sentiment a different object
from the overall tone of the sentence. Dense embeddings also allow
researchers to compare passages that share few words and to use these
representations in retrieval or return models
\citep{lopez2025v6,embeddinggemma2025}. The relevant unit of evaluation
therefore depends on whether a study uses the article, a mention within
it or the link between the article and an issuer.

In CCFinance we apply a publisher filter to a frozen English Fine News
release derived from CC-News, yielding 14,807,287 staged records.
To characterize that selection, we estimate the yield of the same
hostname rule across four pinned source releases. Within the frozen
sample we check publication dates against capture, select canonical
copies, and use schema-driven GLiNER2.5 Small to extract company spans
and attributes \citep{gliner25}. We compare its spans with Stanza,
evaluate dated issuer links and subject assignments against held-out
model judgments, and audit the additional labels and body embeddings.
By holding the annual article requirement fixed, we can measure how
many issuer-years cease to qualify when a study requires each article
to concern the firm. The upstream Fine News revision is publicly
accessible, although reconstructing the exact CCFinance sample also
requires the restricted staging snapshot and derived artifacts.

We trace the consequences of these choices through selected designs
from three peer-reviewed studies. At the publisher-month level, we
compare the equity-market volatility tracker of \citet{baker2026} with
one built from archived articles on the same calendar. When the unit is
a pair of banks, as in the co-mention design of \citet{kazinnik2026},
we find that connected pairs rose across publishers and fell within
Reuters. For the overnight design of \citet{lopez2026}, a
story must also have been available before the trading cutoff.
Recovering publication times lets us evaluate both portfolio legs on
more sessions, while mean drift remains negative after refitting.

%% file: sections/02_literature_review.tex
\section{Literature Review}
\label{sec:literature}

What constitutes an observation in a financial news corpus depends in
part on how the news was collected. In constructing FNSPID,
\citet{dong2024} begin with the headlines and URLs on Nasdaq's
per-symbol pages, retrieve bodies from the linked pages, and supplement
the collection with earlier news data. Their modeling experiments start
from 15.7 million time-aligned news records, for which they separately
examine the availability of text and stock symbols. Collecting by symbol
gives the researcher a route from the security to the story. Starting
from a web archive reverses that order: the researcher observes a copy
of the text and must establish which, if any, securities it concerns.
A story indexed under two symbols can supply two company observations,
while two captures of the same page can document a revision. Counts of
time-aligned records and archived copies consequently measure different
aspects of news coverage. They also suggest different checks on
completeness. Starting with a list of securities, one can ask which
ones have any retrievable text at all. Starting with archived articles,
one can ask how often a recognized company resolves to a listed issuer.
To learn which firms receive little or no coverage, however, both
approaches need a roster defined independently of the articles found.
Otherwise an issuer with no stories disappears from the denominator
along with the evidence of its absence.

For researchers starting with the archived page, the choice of corpus
also determines which editorial and processing decisions they inherit.
The daily CC-News files preserve fetched responses from news sites
\citep{nagel2016}, while Infini-News extracts article text and metadata
that can be searched together with their capture provenance
\citep{lazzaroni2026}. Fine News further filters the extracted records by
language and quality and selects representatives of duplicate texts
\citep{solovev2026}. For pretraining, filtering and deduplication help
determine which passages a model encounters and how often it encounters
them \citep{penedo2023}. For a financial panel, the same operations
can change the measured volume of coverage. Two publishers carrying the
same wire story may each contribute a publication to a study of media
attention, although counting both gives an additional observation of the
same bank pair in a study of co-mentions. The appropriate treatment of
the second copy depends on which of these quantities the study estimates.

A further distinction arises between finding an organization in the
text and identifying a security that a researcher can follow over time.
GDELT's Global Knowledge Graph records organizations, themes and tone
in monitored news and therefore allows researchers to follow how events
are reported \citep{gdelt}. FiNER-ORD makes recognition itself amenable
to evaluation: \citet{shah2024} provide manual annotations for people,
organizations and locations in 201 financial news articles. Those labels
let a researcher ask whether an extractor found the right words in a
passage. In a financial application, a correctly extracted organization
might be an exchange, a regulator or a subsidiary with a listed parent,
each of which has a different implication for whether the mention can
enter an issuer panel. \citet{schwenkler2020} make this additional step
explicit when they recognize organizations in news and link firms to
construct a network. A missed name can remove a genuine pair, while an
incorrect assignment can attribute that pair to a different firm.

Even a correct assignment leaves open what the text says about the firm.
Of the 10,753 human-annotated headlines in SEntFiN, 2,847 contain several
entities, and the sentiments expressed toward them can differ
\citep{sinha2022sentfin}. Consider a headline announcing the acquisition
of one company by another. The buyer and the target should both be
retrievable by name, though an investor interpreting the terms of the
transaction may draw different conclusions for each security.
\citet{heston2017} can study such firm-level questions by matching news
text to vendor firm tags and sentiment scores. For archived text, the
corresponding decisions can be examined separately by asking whether
the name was linked to the right issuer, whether the story concerned
that issuer, and whether the sentiment assigned to it agrees with a
judgment made in context.

The financial application determines which of these decisions enters
the estimate. In the volatility tracker of \citet{baker2026}, articles
that meet a dictionary rule contribute to a publisher's monthly share,
so the other articles the publisher carried that month also enter the
calculation through the denominator. In the bank networks of
\citet{kazinnik2026}, a connection requires evidence about two
institutions in the same story. The overnight return design of
\citet{lopez2026} adds a time requirement, since a headline can inform
a trading decision only if it was available before the position was
taken. For a captured article body, that requirement concerns the
version containing the information used by the model. We therefore
begin the construction of CCFinance with the archived copy and carry
its identity forward as we assign issuers, evaluate their roles in
the story and establish which observations can enter each design.

%% file: sections/03_methodology.tex
\section{Methodology}
\label{sec:method}

To identify the archived copies from which a financial observation can
be constructed, we begin with the English-language \texttt{eng\_Latn}
shards of Fine News. We retain a record when its hostname matches an
entry on our 115-entry publisher list or a dot-delimited subdomain of
that entry. The list includes general-interest publishers with financial
coverage. For each match we preserve the article ID, body, source URL
and available capture and publisher metadata. The upstream revision
and a separate checksum-verified staging revision identify 14,807,287
distinct IDs in 168 staged Parquet shards (see Appendix
Table~\ref{tab:app-source} for the source comparison).

To see how the choice of source affects the yield of this filter, we
apply the same hostname rule retrospectively to four pinned releases,
including FineWeb-HQ \citep{messmer2025}. Within each release we make
four independent draws with replacement, sampling a file uniformly and
then a row group uniformly within that file. We use inverse-probability
weights to estimate the publisher-matched and English totals, whose ratio
gives the estimated publisher yield. The two news collections have the
higher point estimates (Appendix Table~\ref{tab:app-source}). Because
the releases differ in their English-language coverage and treatment of
duplicates, the yields describe physical rows within each release, while
estimating how many unique stories a source adds would require matching
records across releases.

Within the chosen release, an article is usable if it is the canonical
copy and its reported publication date is consistent with capture.
Although Fine News retains representatives of upstream duplicate clusters,
the staged records still contain exact-text and normalized-URL repeats.
We select the earliest capture separately within each text group and
each normalized-URL group, breaking ties by ID and recording the text
match when both comparisons identify a duplicate. A copy must win both
comparisons before it can qualify, even if an earlier copy that takes
precedence subsequently fails the date check. Because publishers can
revise the text at the same URL, the canonical ID identifies the
particular body retained.

For the date check we allow capture up to seven days after reported
publication or, to accommodate time zones, up to one calendar day before
it. A page dated Monday and fetched Thursday can therefore count toward
Monday's news in a day-level corpus, although Thursday is the first
date on which we observe that version for a trading signal. In the
frozen corpus, 237,193 of the 11,260,839 usable copies were captured
between two and seven days after reported publication. There is also
direct evidence that a later visit can return a different body:
among 26,897 repeated normalized URLs with observed text hashes,
23,779 had more than one hash. A change in hash alone does not tell
us whether a publisher added financially relevant information, but
it does explain why we retain the capture time of the particular text
from which the signal is computed.

To determine which firms that version names, we use GLiNER2.5 Small
\citep{gliner25} to extract candidate company spans, retain their text
and offsets, and resolve them against a table of dated listings and
aliases. The table combines SEC registrant identifiers and current and
former names with listing, alias and subsidiary evidence from Wikidata.
Where validity dates are known, we evaluate alias evidence as of the
article date. An exchange ticker or Reuters code can anchor an identity
and help resolve later mentions within the article, while ambiguous
candidates are retained and the linker can abstain. In the anchor
audit, for example, the issuer associated with a ticker
in an older Sprint story differs from the issuer to which that ticker
was later assigned.

\citet{schwenkler2020} likewise identify organizations with Stanford
CoreNLP before linking firms in a news-derived network. To examine
company recognition before the linking decision, we compare exact
company spans from GLiNER2.5 Small and Stanza, a separate Stanford
toolkit, in leads from our publisher-filtered sample (Appendix
Table~\ref{tab:app-spans}). FiNER-ORD also documents the difficulty of
transferring named-entity recognition to financial text \citep{shah2024}.

Even when the issuer's identity is clear, its role in a story can vary.
A dispute involving Gap and a Tapestry brand can concern both firms,
while an article about Tesla may mention Nvidia in passing. We retain
each accepted article--issuer pair with the evidence for its link,
and count the article as focused on an issuer when it contains at least
three extracted mentions of that issuer, with no further relevance
threshold. A usable story that names one linked firm three times and
another once thus counts once in the corpus, toward usable coverage for
both firms, and toward focused coverage for the first.

To describe the text at each of these levels, GLiNER uses a joint schema
on 384-word windows with 64 words of overlap, recording article form,
local relevance and investor-facing sentiment for extracted mentions,
and event scores for individual windows. For the overnight probe,
EmbeddingGemma encodes up to 2,048 tokens per story in a fixed
768-coordinate body vector \citep{embeddinggemma2025}. We check vector
norms and alignment with the frozen article IDs and examine
company-specific representations in a separate exploratory retrieval task
(Appendix Table~\ref{tab:app-vectors}).

The distinction between a body and an issuer also applies to its
representation. If a story names both parties to a transaction, the
stored body vector is the same whichever party a researcher queries.
Conditioning an embedding on the sentences that mention one party
could help distinguish the two accounts, provided those sentences
contain different information. The exploratory retrieval task tests
that condition by asking each company's representation to retrieve
its own account after the names have been masked. Only 16 held-out
stories judged eligible by the primary model fit every encoder,
below the planned minimum of 60, and two companies can still share
the same selected sentence. We therefore use the body vector for the
overnight exercise and retain the article--issuer links as a separate
record of which firms the story concerns or mentions.

%% file: sections/04_empirical_design_and_data.tex
\section{Data and Empirical Design}
\label{sec:design}

We count coverage first by article and then by issuer, since a single
canonical copy may supply news about several firms. Panel~A of
Table~\ref{tab:coverage} counts each distinct article ID once per row,
while Panel~B follows issuers across the nine complete calendar years
from 2017 through 2025. For a given year an issuer qualifies with
either a dated listing or an
undated current listing supported by public-float evidence in that year
or the preceding year. Qualifying issuers remain in the denominator
even when no articles could be linked to them.

Because we assign news to the publication year but observe it at
capture, a year with many archived pages need not supply many eligible
stories about firms in that year. The 2017 capture cohort illustrates
the distinction. Of the 3,659,035 staged records captured that year,
2,886,747 were fetched more than seven days after reported publication,
including over a million late captures on each of the UK and Indian
Reuters hosts (Appendix Table~\ref{tab:app-coverage}). Counting all
those pages in 2017 would conflate the date on which a crawler visited
a publisher with the date on which the publisher reported the news.
Under the publication-year rule, 854 of 3,240 eligible issuer-years
clear the 50-usable-article minimum in 2017, compared with 1,844 of
3,607 in 2020. The fact that a firm qualifies in one year therefore
does not establish continuous coverage over the study period.

Of 34,781 eligible issuer-years, 33,024 have at least one usable article.
At an annual minimum of 50 articles, 13,918 issuer-years qualify on
usable links, while 8,641 meet the same minimum when each story must
mention the issuer at least three times. The cost of that restriction
also varies with firm size. In the lowest annual quintile of measured
public float, 329 of 6,533 issuer-years (5.0\%) clear the usable-article
minimum and 105 (1.6\%) clear the focused minimum. In the highest
quintile the corresponding counts are 5,790 of 6,526 (88.7\%) and
4,561 (69.9\%). Appendix Table~\ref{tab:size-year} gives the
intermediate quintiles and the breakdown by year. Among smaller firms,
even the initial requirement selects a narrow group: roughly one in
twenty issuer-years in the lowest quintile reaches the 50-article minimum.
Requiring subject focus then retains 105 of the 329 that cleared the
first screen, compared with 4,561 of 5,790 in the highest quintile.
For an analysis that estimates one coefficient across the qualifying
firms, changing the definition of a relevant story consequently changes
the composition of the sample on which that coefficient is estimated.

\begin{table}[t]
  \centering
  \small
  \begin{tabular}{@{}lr@{}}
    \toprule
    Panel A. Article requirements & Article IDs \\
    \midrule
    Staged & 14,807,287 \\
    Consistently dated & 11,287,476 \\
    Usable canonical copy & 11,260,839 \\
    Usable with an accepted issuer link & 5,220,290 \\
    Usable with a focused issuer link & 2,547,597 \\
    \bottomrule
  \end{tabular}

  \smallskip
  \begin{tabular}{@{}lrrr@{}}
    \toprule
    Panel B. Annual minimum & Raw & Usable & Focused \\
    \midrule
    1 article & 33,109 & 33,024 & 31,745 \\
    10 articles & 26,930 & 26,434 & 22,243 \\
    50 articles & 14,539 & 13,918 & 8,641 \\
    \bottomrule
  \end{tabular}
  \caption{Distinct staged IDs surviving the article-level rules
    (Panel~A) and issuer-years meeting each annual article minimum
    during 2017--2025 (34,781 eligible issuer-years in Panel~B). An
    article can count for several issuers
    and qualifies as focused for each linked issuer it mentions at least
  three times.}
  \label{tab:coverage}
\end{table}

To understand which stories enter an issuer's coverage, we examine
first whether the extractor recognizes a company name and then whether
the linker assigns that name to the right issuer. Of the 712
model-reference company spans in 250 sampled leads, GLiNER recovered 412
and Stanza 421 under exact matching, while allowing overlapping boundaries
brings their company-span F1 scores within half a percentage point
(Appendix Table~\ref{tab:app-spans}). Exact scoring requires the first
and last characters of a predicted name to agree with the reference.
For example, Stanza attached a possessive to the name of Credit Suisse
Group AG in a lead that also named UBS Group AG. A reader could
recognize both banks, yet the extra characters cost Stanza the pair
under exact scoring. Conversely, an organization correctly recognized
under Stanza's broader label counts as a false positive when it is
outside the business-only reference. The comparison therefore describes
what each extractor supplies to a company-linking task, whose next
step is to decide whether the name identifies an eligible issuer.

For the identity comparison, we sample extracted spans independently of
the linker's decisions and ask model judges to identify the issuer on
the article date from the marked text and neutral search results, with
the predictions withheld. On held-out extracted spans,
the 475 accepted links have 96.8\% estimated precision (95\% interval
94.0--99.6\%), and recall among reference-linkable mentions is 70.8\%
(64.6--77.1\%). On the same 886 held-out mentions, weighted issuer-link
F1 is 81.81\% with the full linker and 59.46\% when a normalized name
must identify one dated issuer. Restricting learned aliases and their
frequencies to prior calendar years yields 78.57\%
(Appendix Table~\ref{tab:app-link-baselines}). The two time restrictions
have different purposes. The listing date determines whether a candidate
could have been the issuer at the time of the article, while the prior
window determines which earlier uses of a name can help resolve it.
Of the 475 production assignments, the prior-year replay keeps 464,
changes the identity on one and abstains on ten. Thus a historical user
can lose a link even when both versions of the linker consult the same
dated listing.

Having established which issuer a name identifies, we ask whether the
article concerns that issuer. This comparison uses article-level
judgments and is conditional on GLiNER detecting at least one company
in the article. Among the relevance, position and mention-count rules
compared on 217 development articles, the three-mention rule achieved
the highest subject F1. On 183 separate test articles, it retained 37
of 65 judged subject pairs. Relative to counting all accepted links,
requiring three mentions raised subject precision from 31.9\% to 66.1\%
(54.8--77.3\%) but reduced recall from 92.3\% to 56.9\%
(44.3--69.5\%). Of the 188 accepted test pairs, the judges identified
60 as subjects. Requiring three mentions retained 56 pairs, of which
37 were subjects. The increase in precision is useful when a study
needs articles principally about the firm, although it excludes 23 subject pairs
that the broader rule had retained. A researcher willing to admit
more incidental
mentions can make a different choice on the same records: a two-mention
minimum recovers 76.9\% of judged subjects at 55.6\% precision on the
held-out articles (Appendix Table~\ref{tab:app-focus-rules}).

Because the production panel counts named mentions, a story can fall
short of three even when it continues to discuss the firm using a
pronoun or a description. When we also count nominal and pronominal
mentions in Stanford CoreNLP chains anchored to a single linked issuer
on the same test articles \citep{manning2014corenlp}, twelve additional
article--issuer pairs qualify, seven of which are judged subjects.
Recall rises from 56.9\% to 67.7\% (paired gain of 10.77 percentage
points, 95\% interval [3.92, 19.61]), while precision falls from
66.1\% to 64.7\%. A separate GPT-6.1 Terra review attributed 28 of
the 42 added references in those pairs to the anchored issuer. Seven
of the twelve pairs still qualify using only those confirmed references,
and six of those seven are judged subjects.

Once a story is assigned to a firm, its tone can still differ from
what the news implies for that firm's investors. On the 358 of 400
marked mentions that the primary judge labeled positive, negative or
neutral, GLiNER's company-specific sentiment achieved 46.79\% weighted
macro-F1, compared with 40.84\% when each company received the
Stanford sentiment of its sentence. The paired 95\% interval for the
5.95-point difference spans zero [$-1.33$, $+13.98$]. A blinded
GPT-6.1 Terra review marked 23 of 40 sampled targets as invalid
company spans. We therefore treat issuer sentiment as exploratory.

Similar questions arise when researchers select articles by form or
event. To assess how those choices affect sample composition, blinded
model judges label 400 sampled units in each of three confirmation
frames covering article form, local mention attributes and event
families in input windows. The weighted
press-release share among usable articles is 42.1\% under the
classifier and 4.2\% under the primary judge, with 14 reference-positive
sampled articles. A filter on predicted transactions would miss many
windows the judge labels as transactional, since recall against those
labels is 23.6\%. Appendix
Table~\ref{tab:app-metadata} reports estimates by class. These errors
would enter two common designs differently. Removing predicted press
releases would discard articles the reference judge regarded as other
forms of news, whereas retaining only predicted transactions would omit
many windows containing transactions according to that judge. The unit
matters here as well: a missed window establishes a local omission,
while exclusion of an entire article depends on what was detected in
its other windows.

The unit on which these decisions operate changes across the three
financial applications. For the equity-market volatility tracker we
divide the number of articles containing the required groups of words
by all eligible articles from that publisher in the same month. For a
bank network we also need to know which institutions appear together,
since each eligible article contributes one observation for every
unordered pair of mapped banks it mentions in the declared week. A
three-bank story thus supplies three pairs, while the weight on each
pair counts the stories in which it appears.

For the overnight exercise, identifying the security also determines
which exchange session to use. We pair timely news from two Indian
publishers with US-listed securities, which trade on Nasdaq in 740
article--security rows of the fixed panel and on the NYSE in 364.
For 270 rows the registry lacks a
listing start date, so eligibility also requires usable pre-2021 news
linked to the issuer. To establish what could have been observed before
the open, we require first publication to follow the preceding exchange
close and precede 09:00 New York time, with capture between publication
and the cutoff, inclusive. If the crawler fetches an overnight story
after 09:00, we observe the body only after the cutoff. We measure
drift from the open to the close because the close-to-open reaction
begins before the article appears. The
rolling probes train on returns realized before each refit, while
the encoder was released after the study period and its pretraining
overlap with the archived stories is unmeasured. Each active long or
short leg requires at least two
securities, so a session with predictions contributes zero to the
portfolio mean when neither leg qualifies,
while a day without predictions falls outside the evaluation.

%% file: sections/05_results.tex
\begin{table*}[t]
  \centering
  \small
  \begin{tabular}{@{}L{0.15\textwidth}L{0.22\textwidth}
    L{0.25\textwidth}L{0.29\textwidth}@{}}
    \toprule
    Exercise & Evaluated unit and support & Baseline and reference &
    Sensitivity \\
    \midrule
    Volatility
    & 49 common months, 675,234 eligible articles, four publishers
    & VIX $R^2$ 0.515 vs. 0.647, gap $-0.132$ [$-0.215$, $+0.110$]
    & Gap $-0.041$ after dropping March--April 2020 from both fits \\
    Bank network
    & Distinct unordered bank pairs in each earnings week, 33 of 34
    reference nodes mapped
    & 177 to 242 pairs vs. 210 to 172 in the July 2023 manuscript
    & Reuters only, 50 to 35. Matched mean increases $+56.3$ [5, 83]
    on six weeks and
    $+19.1$ [$-25$, 59] on fourteen weeks \\
    Overnight drift
    & 277 evaluated sessions, 42 with both legs and 123 with neither
    & $-10.15$ bp [$-25.2$, $+4.2$] vs. $+11$ bp for the reference
    MiniLM headline sample
    & Recovered clocks give 340 evaluated sessions, 110 with both
    legs, and $-7.73$ bp after refitting \\
    \bottomrule
  \end{tabular}
  \caption{For volatility, the paired six-month block interval measures
    uncertainty in the difference in contemporaneous fit. For banks, the
    ranges cover the central 95\% of matched-story draws, with 682 stories
    per week in the six-week analysis and 158 in the fourteen-week analysis.
    Drift is measured in gross basis points per evaluated session, with a
  21-session block interval conditional on the saved predictions.}
  \label{tab:studies}
\end{table*}

\section{Results}
\label{sec:results}

Table~\ref{tab:studies} compares the sensitivity of the three estimates
to the decisions that determine their samples. We vary the months and
publishers that contribute to the volatility tracker, the stories in
which banks can appear together, and the evidence that a firm's news
was available before the trading cutoff.

\subsection{Publisher-Month Volatility}

The volatility benchmark asks how much of the same monthly VIX variation
each archive's tracker explains. We first reproduce the six monthly columns of
\citet{baker2025ms,baker2026} to their displayed precision using the
authors' archived data \citep{baker2025archive}. The CCFinance tracker
uses 675,234 consistently dated canonical articles from Benzinga,
Forbes, MarketWatch and the \emph{Wall Street Journal}, the only one of
the four in the authors' eleven-newspaper basket. A hit requires terms
from all three dictionary groups for the economy, equity markets and
volatility across the title and body. We divide each publisher's hit
shares by their dispersion over 40 calibration months through December
2022, average the standardized shares, and rescale the average to the
calibration-period mean VIX. From September 2019 through September
2023, each tracker enters a separate regression with an intercept and
the same monthly VIX outcomes from FRED. The CCFinance tracker has an
$R^2$ of 0.515 against 0.647 for the authors' tracker, a difference of
$-0.132$ (paired 95\% interval $[-0.215, 0.110]$).

Omitting March and April 2020 from both regressions reduces the
variation each tracker explains and narrows the VIX fit gap to
$-0.041$. With calibration held fixed, omitting Forbes widens the gap
on either calendar, taking it
from $-0.132$ to $-0.360$ on the full sample. When we retain only the
\emph{Wall Street Journal}, the publisher common to both baskets, its
tracker explains a smaller share of monthly VIX variation
($R^2=0.110$) and has a correlation of 0.397 with the authors'
eleven-paper composite. These comparisons help interpret the fit gap
because a common calendar equalizes the volatility outcomes but leaves
each tracker's news denominator intact. If a publisher expands its
market coverage in a volatile month, the dictionary-hit share depends
both on how many of the new stories discuss volatility and on how much
other news the publisher carries. Standardizing each publisher's share
puts its variation on a common scale while preserving that dependence
on the mix of stories. The single-newspaper comparison uses the
reference composite because the supplied monthly benchmark aggregates
all eleven papers (Appendix Table~\ref{tab:app-emv}).

\subsection{Bank Pairs in Earnings Weeks}

Publisher choice also determines which banks appear together, since
a pair enters the network only when a retained story names both
institutions. For the weeks of 13--19 January and 13--19 April 2020
in the July 2023 design of \citet{kazinnik2023ms,kazinnik2026}, we
replace the Factiva tags with CCFinance links for 33 of the 34 reference
nodes, in some cases through a foreign group's listed parent. The six
largest institutions form a complete core in the reconstructed networks,
and the number of connected pairs rises from 177 to 242, whereas the
July 2023 manuscript reports a decline from 210 to 172. Restricting
CCFinance to Reuters changes both the publisher mix and the number of
available stories, and within that sample the count falls from 50 to
35. The increase persists when we exclude Nasdaq alone (162 to 230)
and when we retain all publishers but use only direct bank mappings
(77 to 133).

To see whether the increase survives at fixed publisher quotas, we
collapse syndicated copies and match the number of stories drawn from
each publisher. Over six weeks with 682 matched stories per week, the
mean January--April 2020 increase is 56.3 connected pairs, with a
central 95\% subsampling range of [5, 83]. On the same six weeks,
removing Truist to obtain a roster usable before its 2019 merger reduces
the increase to 45.4 pairs. Requiring the same publisher quotas across
fourteen weeks then leaves 158 stories per week and excludes Nasdaq,
which supplied 271 of the original 682. The mean increase falls to
19.1 pairs [$-25$, 59], close to the 19.0 added from July to October
2020 under the same quotas. These ranges capture variation across
draws of stories from the
chosen weeks, while the six placebo periods show how the network
changed at other dates (Table~\ref{tab:app-banks}). A comparison over
more weeks thus asks the archive to supply a smaller set of publishers
consistently for longer. The January--April increase on that common
sample is about as large as the subsequent July--October change, even
though the unmatched January and April networks differ by 65 pairs.

For a network, a bank can contribute evidence whenever it appears
alongside another bank in the same week. In the overnight design the
timing requirement is tighter in two respects: each article must enter
before its own trading cutoff, and several securities must qualify on
the same date before their returns can be combined in a portfolio.

\subsection{Issuer-Session Drift}

For the overnight exercise, the coverage question is whether enough
securities have timely news in the same session to form portfolio legs.
We apply the selected design of \citet{lopez2025v6,lopez2026} to 996
captured bodies from Business Standard and Mint Lounge linked to 171
issuers, using EmbeddingGemma vectors and Yahoo price histories in place
of the authors' RavenPack-matched headlines and CRSP prices. Following
their archive, we fit separate rolling logistic regressions on prior
return signs in six-month windows \citep{lopez2026archive}. Within
each session the earliest eligible predicted article supplies
the signal for its security. We weight securities equally within each
qualifying leg and sum the signed returns on one unit of notional per
active leg, assigning zero to an inactive leg. Although the probes
train on earlier returns, the issuer linker estimates alias frequencies
across the full archive, so a historical link can draw on later evidence.

Of 32 months, twelve had fewer than 100 training rows, and in 16 of
the 20 fitted months the baseline solver warned at the ten-iteration
cap. The fixed-sample probes evaluate 277 of 670 possible sessions
from October 2021 through May 2024. Only 42 of the evaluated sessions
have both a long and a short portfolio, while 123 have neither. Mean
open-to-close drift before costs is $-10.15$ basis points per evaluated
session, with a 21-session block interval of [$-25.2$, $+4.2$] basis
points conditional on the saved predictions. The denominator includes
every evaluated session because a portfolio that forms neither leg
deploys no capital and earns zero that day. Conditional on at least
one active leg, the same gross returns average $-18.26$ basis points
on 154 dates. The two means describe, respectively, the return per
evaluated session and the return on dates on which the rule actually
took a position (Appendix Table~\ref{tab:overnight-support}).

Recovering first-publication times from the historical Business Standard
format adds 667 distinct firm-sessions. With those additional
observations the number of evaluated sessions
rises from 277 to 340, of which 110 have both legs, up from 42, and
74 have neither. Because the recovered observations also enter the
training windows, we refit the probes on the expanded panel, where mean
drift remains negative at $-7.73$ basis points per evaluated session.
On the
277 dates evaluated by both models, recovery changes mean gross drift
by $+2.45$ bp (paired 95\% interval $[-6.80,+13.74]$). Whether this
gain survives a trading cost depends on how many legs actually trade
on those dates. When each active leg incurs a cost of five basis points
one way at both entry and exit, the paired change on the common dates is
$-1.27$ bp $[-11.19,+9.25]$. Additional positions incur additional
costs even when the calendar is held fixed, and the intervals accommodate
either sign for both the gross and net changes (Appendix
Table~\ref{tab:overnight-support}).

Table~8 of
\citet{lopez2025v6} reports $+11$ basis points of mean daily drift for
MiniLM headlines in a broader RavenPack and CRSP sample, with different
texts, issuer coverage, representation and prices. On the fixed
CCFinance sample, the paired title and body comparisons leave substantial
uncertainty about which representation produces a higher mean drift
(Appendix~\ref{sec:app-studies}).

The stability of the fitted probe matters here because a change in
one security's signal can determine whether its leg has the required
two members. Allowing the baseline solver up to 1,000
iterations at the original regularization strength removes the 16
warnings and changes nine of 738 security--session signals. The same
277 dates are evaluated, yet only 39 now support both legs and mean
drift is $-8.73$ basis points. Recovering timestamps gives more
opportunities to form a portfolio, while the signs assigned to eligible
securities and the returns they subsequently earn determine what that
additional coverage is worth. Appendix Table~\ref{tab:app-drift}
reports the cost scenarios for each representation.

%% file: sections/06_conclusion.tex
\section{Conclusion}
\label{sec:conclusion}

An archived article can enter a publisher's monthly news count as soon
as its source, date and status as a distinct copy are established.
For a bank network, the same article contributes a relationship only
when it names two identified banks, and the treatment of syndicated
copies determines how much weight that relationship receives. An
overnight portfolio requires a further judgment about what was
observable before the cutoff and enough eligible securities in the
same session to form a leg. Recording these decisions alongside an
estimate allows a researcher to see both how much coverage a stricter
rule costs and which firms or dates account for the difference.

CCFinance preserves this record for a frozen selection and provides
audits of issuer identity, subject focus and supporting metadata.
Although the upstream archives and their documentation are publicly
accessible under their respective terms, anonymous requests to the
staging snapshot required authentication, and access to the project
repository and frozen outputs is restricted during review. Once the
required inputs are restored, the recipes can regenerate the reported
analyses (Appendix Table~\ref{tab:app-access}). We plan to release the
code and eligible derived records after review, with terms specified
separately for each artifact. For researchers constructing a new panel,
the ledger makes it possible to trace a change in firm coverage or
portfolio support back to the individual stories that qualified.

%% file: appendix.tex
\input{sections/07_appendices}

%% file: sections/07_appendices.tex
\setcounter{table}{0}
\renewcommand{\thetable}{A\arabic{table}}

\section{Corpus lineage and source pilot}

To apply the historical publisher list to English-language Fine News
records, we take the hostname from the metadata, falling back to the URL
when necessary, and include dot-delimited subdomains of the listed hosts.
The replication package identifies both the upstream revision and the
frozen selection of 14,807,287 distinct IDs across 168 shards. Verifying
that a fresh upstream selection reproduces the snapshot still requires
a row-by-row comparison of IDs and bodies.

Within each of four pinned release frames (Table~\ref{tab:app-source}),
the retrospective pilot samples a file and then a row group within it.
We expand each count by the inverse of the product of its two selection
probabilities and take the ratio of the resulting Hansen--Hurwitz totals
to estimate publisher yield, with exploratory 95\% linearized $t$
intervals on three degrees of freedom. The manifests preserve the release
versions, draw probabilities and observed counts. Because filtering and
deduplication differ across releases, these totals count physical rows.
Estimating unique stories, overlap or financial coverage outside the
registry would require separate sampling.

With only four draws, the interval for the Fine News matched total is
wide enough (2.15--18.67 million physical records) to encompass the
14.81 million staged IDs. Those IDs amount to 6.76\% of the
219,192,987 English records in the pinned release, although the exact
selection has yet to be independently replayed.
The Fine News, Infini-News, RefinedWeb and FineWeb-HQ frames contain
168, 49,748, 5,534 and 9,246 files respectively. The four selected
Fine News files contain 437, 519, 71 and 319 row groups. Each sampled
Infini-News file contains one group and each RefinedWeb file contains
175. The FineWeb-HQ files contain 352, 336, 164 and 331 groups.

\begin{table*}[t]
  \centering
  \small
  \setlength{\tabcolsep}{4pt}
  \caption{Estimated publisher yield in four source releases. Within each
    release we make four independent file--row-group draws with replacement
    and divide the estimated number of matching English physical records
  by the estimated number of English physical records in the frame.}
  \label{tab:app-source}
  \begin{tabular}{llrrrl}
    \toprule
    Source & English definition & Sampled English & Matched &
    Est.\ matched & Yield [95\% interval] \\
    \midrule
    Fine News & English shard paths & 16,384 & 799 & 10,408,482 &
    4.50\% [2.36, 6.63] \\
    Infini-News & Native \texttt{eng\_Latn} & 37,632 & 2,575 &
    32,025,275 & 6.84\% [3.37, 10.32] \\
    RefinedWeb & Release scope & 4,000 & 20 & 4,842,250 & 0.50\% [0.16, 0.84] \\
    FineWeb-HQ & Native \texttt{en} & 4,000 & 25 & 18,887,267 &
    0.69\% [0.13, 1.25] \\
    \bottomrule
  \end{tabular}
\end{table*}

Among 11,260,839 usable copies, 237,193 were captured two through seven
days after publication. Among 26,897 repeated normalized URLs with
observed text hashes, 23,779 carried more than one body hash. An
earlier publication date alone cannot establish when a revision appeared.

\paragraph{Historical selection.}
The original filter comprised the 115 entries below, but even subdomains
of a single publisher have distinct capture histories. Among the five largest
observed hosts, the staged and usable shares of the respective totals
were 9.0\% and 11.6\% for nasdaq.com, 8.9\% and 0.2\% for uk.reuters.com,
7.7\% and 0.9\% for in.reuters.com, 6.5\% and 5.9\% for reuters.com,
and 5.7\% and 7.5\% for business-standard.com.

{\small\raggedright accountingtoday.com, afr.com, agriculture.com,
  agweb.com, barrons.com, bdlive.co.za, benzinga.com,
  bitcoinmagazine.com, bloomberg.com, bloombergquint.com,
  business-standard.com, businessdailyafrica.com, businessday.co.za,
  businessinsider.com, businessinsider.com.au, businessinsider.de,
  businessinsider.pk, businesslive.co.za, businesstimes.com.sg,
  businesstimesafrica.net, businessweek.com, businesswire.com, cfo.com,
  channelnewsasia.com, cityam.com, cnbc.com, coindesk.com,
  cointelegraph.com, dailyfx.com, dawn.com,
  economictimes.indiatimes.com, economist.com, efinancialnews.com,
  etf.com, etfdb.com, etftrends.com, etfworld.co.uk, fin24.com,
  finance.yahoo.com, financial-planning.com, financialexpress.com,
  financialpost.com, fnlondon.com, fool.co.uk, fool.com, fool.com.au,
  forbes.com, forexcrunch.com, fortune.cnn.com, fortune.com,
  freightwaves.com, ft.com, globenewswire.com, gurufocus.com,
  handelsblatt.com, ibtimes.co.in, ibtimes.co.uk, ibtimes.com,
  ibtimes.sg, institutionalinvestor.com, investing.com,
  investingnews.com, investmentweek.co.uk, japantimes.co.jp, kitco.com,
  koreabizwire.com, koreaherald.com, koreajoongangdaily.joins.com,
  law360.com, livemint.com, malaymail.com, marketwatch.com, mining.com,
  miningmonthly.com, moneycontrol.com, moneyweb.co.za, morningstar.com,
  morningstar.com.au, morningstar.in, morningstaronline.co.uk,
  nasdaq.com, nikkei.com, nikkeibp.co.jp, nikkeibp.com,
  offshore-energy.biz, oilprice.com, pionline.com, platts.com,
  prnewswire.co.uk, prnewswire.com, proactiveinvestors.co.uk,
  proactiveinvestors.com, proactiveinvestors.com.au,
  professionalpensions.com, qz.com, reuters.com, rigzone.com,
  rttnews.com, scmp.com, seekingalpha.com, spglobal.com,
  straitstimes.com, taipeitimes.com, theedgemalaysia.com,
  theedgemarkets.com, theedgesingapore.com, thefinancialexpress.com.bd,
  themalaymailonline.com, thestreet.com, thinkadvisor.com,
thisismoney.co.uk, upstreamonline.com, worldoil.com, wsj.com, zacks.com.\par}

\section{Extractor and metadata audits}

\paragraph{Exact company spans.}
The microbenchmark takes the 50 lowest seeded ID hashes per
publication year from 2019 through 2023 among consistently dated,
canonical English articles in the staging snapshot. For each lead we
take the first 150 whitespace-delimited
words or the entire body if it is shorter. A GPT-6.2-sol medium primary
pass covered all 250 leads, including the 40-lead pilot, while a separate
GPT-6.2-sol low pass covered the first 12 ranks in each of the five
years. The two passes agreed exactly on 52 of the 60 repeated leads
(span F1 95.68\%). The repeat annotator had access to published pilot
examples, and disagreements were resolved against the guide before
scoring. The resulting model reference contains 712 spans across the
250 leads, of which 73 contain none. The guide includes named businesses
and excludes products, bare tickers and nonbusiness organizations.

Stanza 1.11.0 uses the English \texttt{ontonotes\_charlm} model with
combined tokenization and multiword-token expansion. GLiNER2.5 Small
uses the deployed company schema and cutoff 0.5 \citep{stanza,gliner25}.
Scoring matches Stanza ORG or GLiNER company spans by article ID and
exact half-open Unicode offsets. A correctly recognized noncompany ORG
therefore counts as a false positive for this task.
Table~\ref{tab:app-spans} reports a GLiNER advantage of 10.34
micro-F1 percentage points on this company-specific task, with a
paired 95\% interval of $[5.84,14.56]$ over 2,000 passage bootstraps
within publication year. Differences use unrounded scores. When a
prediction and reference need only share a character, maximum one-to-one
matching gives F1 of 58.13\% for Stanza and 58.54\% for GLiNER.
The paired difference is 0.41 points ($-3.36$ to $+4.40$) over 2,000
year-stratified resamples. Both evaluations retain the business-only reference.
Stanford's published 88.8\% OntoNotes and 92.1\% CoNLL03 scores are
all-class micro-F1 on their benchmark splits \citep{stanza}. In the
local task, a possessive attached to the Credit Suisse Group AG span
cost Stanza its pair with UBS Group AG, while GLiNER missed BofA in a
lead with five stocks and lost five of 15 possible pairs. These are
mention pairs before dated issuer resolution.

\begin{table}[t]
  \centering
  \small
  \setlength{\tabcolsep}{3pt}
  \caption{Exact company-span micro scores (\%). The balanced set has
  250 leads and 712 model-reference spans.}
  \label{tab:app-spans}
  \begin{tabular}{lrrrrrr}
    \toprule
    System & TP & FP & FN & P & R & F1 \\
    \midrule
    Stanza ORG & 421 & 835 & 291 & 33.52 & 59.13 & 42.78 \\
    GLiNER company & 412 & 427 & 300 & 49.11 & 57.87 & 53.13 \\
    \bottomrule
  \end{tabular}
\end{table}

\paragraph{Article, mention and window metadata.}
GLiNER2.5 Small processes 384-word windows with a 64-word overlap.
Each window receives a softmax distribution over seven article forms
and an independent score for each of six event families. The article
inherits the form predicted with the highest confidence across its
windows. Event families are evaluated per window at a cutoff of 0.5,
so a window can have several families or none. The extractor returns
body offsets for company, person, public-body and place spans. Each
company span receives separate softmax scores for local relevance and
investor-facing sentiment. Within an article--issuer pair, the linker
takes the largest substantive-relevance score and averages the sentiment
vectors using mention link scores as weights. The mention-level audit below
evaluates local labels before that issuer-level aggregation.

The seed-46 confirmation excludes seed-45 development articles and
samples 400 units from each usable-corpus frame. For form and sentiment,
half the allocation follows predicted class size and half balances
classes subject to availability. Events split equally between windows
with all scores below 0.5 and positive windows stratified by top family.
The GPT-6 Luna low-effort judge sees the article, marked mention in
local context or event window, with predictions withheld. A second Luna
call and a GPT-6.1 Terra low-effort call independently label the same
100 random units per frame. Estimates use inverse-inclusion weights
and article-cluster multiplier intervals holding stratum totals fixed
(Table~\ref{tab:app-metadata}).

The press-release discrepancy persists using usable article--issuer links
($-37.3$ points, 95\% interval $[-47.0,-27.6]$). Counting three
insufficient-context articles as releases would raise the weighted
reference prevalence from 4.18\% to 4.73\%. Sentiment labels have opposite
positive and negative polarity on 7.4\% of weighted extracted mentions.
Weighted sentiment agreement is 82.3\% on the Luna repeat and 57.2\%
between Luna and Terra. The primary judge found at least one event
family in 160 of 200 sampled windows with all predictions below threshold.
The separate 10,000-article extraction benchmark ran at 106.1 articles
per second with 7.22 GiB peak reserved CUDA memory. Synthetic examples
guided development and contribute zero weight to confirmation estimates.

\paragraph{Sentence and company sentiment.}
The primary judge assigns positive, negative or neutral sentiment to
358 of 400 marked mentions. We compare GLiNER's company label with
the root sentiment of the containing sentence from Stanford CoreNLP
4.5.10, collapsing the two intensities in each direction. Mixed and
unassessable predictions miss the judged class. One Stanford sentence
exceeds the 200-token parsing limit. Weighted three-class macro-F1 is
46.79\% for GLiNER and 40.84\% for Stanford, a paired difference of
5.95 points [$-1.33$, $+13.98$]. Polarity reversals occur on 8.38\%
and 10.73\% of directionally judged units, respectively. A fresh
GPT-6.1 Terra pass agreed with the primary judge on 10 of 40 blinded
cases. It marked 23 targets as invalid company spans, including a
product and a political party. Both span validity and the financial
interpretation of the surrounding text affect this comparison.

\begin{table*}[t]
  \centering
  \small
  \caption{Weighted metadata validation against blinded model judgments.
    The frames each have 400 sampled units (whole articles, extracted spans
    and input windows). $n$ counts unweighted reference positives. Precision
    and recall have 95\% article-cluster intervals. Rare reference classes
    have limited support, including three insufficient-context articles,
    nine roundups and seven invalid spans for relevance. A dash marks
  an empty denominator.}
  \label{tab:app-metadata}
  \begin{tabular}{llrll}
    \toprule
    Task & Class & $n$ & Precision [95\% interval] & Recall [95\% interval] \\
    \midrule
    Article form & Analysis & 44 & .40 [.27,.53] & .53 [.39,.67] \\
    & Insufficient context & 3 & -- & .00 [.00,.00] \\
    & Interview & 15 & .18 [.08,.32] & .31 [.14,.55] \\
    & Opinion & 42 & .18 [.10,.28] & .34 [.20,.49] \\
    & Other & 13 & .14 [.04,.27] & .02 [.01,.07] \\
    & Press release & 14 & .08 [.04,.14] & .80 [.58,.94] \\
    & Reporting & 260 & .86 [.77,.93] & .26 [.23,.28] \\
    & Roundup & 9 & .12 [.04,.26] & .16 [.05,.41] \\
    Local relevance & Incidental & 103 & .41 [.20,.65] & .08 [.04,.15] \\
    & Invalid span & 7 & -- & .00 [.00,.00] \\
    & Substantive & 283 & .71 [.65,.77] & .67 [.61,.73] \\
    & Uncertain & 7 & .00 [.00,.00] & .00 [.00,.00] \\
    Sentiment & Invalid span & 6 & -- & .00 [.00,.00] \\
    & Mixed & 21 & .02 [.00,.09] & .01 [.00,.02] \\
    & Negative & 92 & .50 [.40,.60] & .56 [.47,.65] \\
    & Neutral & 157 & .46 [.35,.58] & .19 [.15,.23] \\
    & Not assessable & 15 & .04 [.00,.11] & .06 [.01,.19] \\
    & Positive & 109 & .41 [.33,.49] & .73 [.66,.79] \\
    Event family & Performance & 150 & .83 [.70,.93] & .21 [.17,.24] \\
    & Financing & 55 & .60 [.43,.77] & .17 [.11,.25] \\
    & Transactions & 60 & .65 [.49,.79] & .24 [.16,.33] \\
    & Operations & 96 & .66 [.49,.80] & .09 [.06,.12] \\
    & Legal and reputation & 93 & .81 [.66,.92] & .02 [.02,.03] \\
    & Markets and macroeconomics & 126 & .87 [.75,.95] & .10 [.08,.12] \\
    \bottomrule
  \end{tabular}
\end{table*}

\section{Issuer identity, focus and coverage}

The seed-43 confirmation samples extracted spans by alias frequency
and the number of issuers sharing a key, and draws a separate uniform
sample of articles containing at least one extracted company span.
Both draws are independent of the linker's predictions. For the span
judgments, the models see the marked text and neutral issuer search
results with the candidate link withheld, while the article judges
receive up to 12,000 body characters. We count a link as correct when
it names the judged issuer or the same listed security on the article
date, and measure unlinked precision by the share of abstentions judged
to have no eligible issuer. Weighted estimates use linearized intervals
(Table~\ref{tab:app-links}).
Cohen's $\kappa$ is 0.935 for the repeated judge on 164 mentions and
0.881 for the independent audit judge on 339.

\paragraph{Selecting a subject rule.}
We compared mention counts, the position of an issuer's first mention
as a fraction of body length, and its highest substantive-relevance
score. Across the grid of six count cutoffs, four position cutoffs and
19 relevance cutoffs, three mentions with no further restriction gave
the highest development F1. On held-out articles, counting two mentions
recovered more subjects and gave a higher F1 than counting three
(Table~\ref{tab:app-focus-rules}). A development-selected relevance
threshold of 0.40 yielded 41.7\% subject precision and 53.8\% recall
on the held-out articles.

Because the judge reads only the first 12,000 characters, some of the
evidence available to the full-body rule can fall outside the judged
text. Seven development and five held-out bodies exceeded that limit.
In one Schlumberger earnings-call
transcript, the third linked mentions of Goldman Sachs, Citigroup and
Bank of America all occurred after character 20,000. The full-body
rule selected all three banks, whereas the judge identified Schlumberger
as the subject from the visible prefix.

\begin{table*}[t]
  \centering
  \small
  \caption{Precision, recall and F1 (\%) for identifying an issuer as an
    article subject. The 217 development and 183 held-out articles contain
    80 and 65 judged subject pairs. Each row varies one rule applied to
    accepted issuer links. The position and relevance rows show the best
  development F1 for each of those measures used alone.}
  \label{tab:app-focus-rules}
  \begin{tabular}{lrrrrrr}
    \toprule
    & \multicolumn{3}{c}{Development} & \multicolumn{3}{c}{Held out} \\
    Rule & P & R & F1 & P & R & F1 \\
    \midrule
    Every link & 27.9 & 83.8 & 41.9 & 31.9 & 92.3 & 47.4 \\
    At least two mentions & 53.8 & 71.3 & 61.3 & 55.6 & 76.9 & 64.5 \\
    At least three mentions & 78.8 & 65.0 & 71.2 & 66.1 & 56.9 & 61.2 \\
    At least four mentions & 88.2 & 56.3 & 68.7 & 75.6 & 52.3 & 61.8 \\
    First mention in first quarter of body & 58.1 & 67.5 & 62.4 &
    54.9 & 69.2 & 61.2 \\
    Substantive relevance $\geq 0.40$ & 48.1 & 63.8 & 54.8 & 41.7 &
    53.8 & 47.0 \\
    \bottomrule
  \end{tabular}
\end{table*}

Against production, the prior-year replay retains 464 issuer identities,
changes one and abstains on ten previously assigned mentions. All 411
production abstentions persist. Registry validity describes the recorded
effective date, while the prior window controls when frequency evidence
could enter the decision. Explicit anchors can resolve a mention directly.
Otherwise the leading candidate needs at least 75\% of the evidence,
with a separate operating score of 0.5. Ambiguous mentions retain their
candidate lists. Frequency strata use fewer than ten, ten through 999,
and at least 1,000 occurrences, crossed with zero, one or multiple
matching issuers.

The anchor audit samples explicit listing anchors outside the seed-43
articles. The decorated-name audit samples newly accepted article--issuer
pairs after normalizing ticker and stock forms. Of 300 newly accepted
pairs, 297 had a trailing ticker and three used the word stock. Six
exchange-prefix cases were checked separately. Normalization adds
49,852 accepted pairs, including 47,808 on usable articles. Agreement
on identity leaves the subject role to be evaluated separately. Short
keys with noncompany meanings and subsidiaries with competing issuer
candidates can produce errors or abstentions. Dated records constrain
valid links, while full-corpus alias priors can use anchors observed later.

\begin{table}[t]
  \centering
  \small
  \begin{tabular}{lrrr}
    \toprule
    Issuer assignment & Precision & Recall & F1 \\
    \midrule
    Unique dated name & 99.12 & 42.47 & 59.46 \\
    WikiDict and crosswalk & 99.78 & 20.42 & 33.91 \\
    Prior-year CCFinance & 96.66 & 66.18 & 78.57 \\
    Full CCFinance & 96.81 & 70.83 & 81.81 \\
    \bottomrule
  \end{tabular}
  \caption{Weighted issuer-link scores (\%) on 886 common extracted
    mentions. Of these, 578 have a model-adjudicated eligible issuer.
    The name rule requires a unique dated issuer. Stanford's WikiDict
    returned a page for 464 mentions, of which 97 matched canonical
    English Wikipedia sitelinks of CIK-bearing Wikidata items and 49
    yielded a unique dated issuer. The prior-year replay fixes learned
    alias evidence before the calendar year and retains the frozen
    registry. Full minus prior-year F1 is 3.24 percentage points
    [1.22, 6.23] in 2,000 paired article-level multiplier draws
  reweighted within sampling strata.}
  \label{tab:app-link-baselines}
\end{table}

\begin{table*}[t]
  \centering
  \small
  \caption{Link and focus diagnostics. Denominators are unweighted counts
    of evaluated predictions or reference positives for the named measure.
    Span estimates use inverse sampling weights. Article estimates use
    the uniform sample of 183 held-out articles with a detected company. The
    anchor and decorated-name rows
  give raw proportions from distinct blinded model audits.}
  \label{tab:app-links}
  \begin{tabular}{lll}
    \toprule
    Measure & Denominator & Estimate [95\% interval] \\
    \midrule
    Extracted-span precision & 475 accepted links & 96.8\% [94.0,99.6] \\
    Extracted-span recall & 578 judged linkable spans & 70.8\% [64.6,77.1] \\
    Unlinked precision & 411 unlinked spans & 84.8\% [80.6,88.9] \\
    Precision with several candidates & 225 accepted spans & 92.0\%
    [86.4,97.6] \\
    Article recall, all mentioned issuers & 202 reference pairs &
    68.8\% [61.0,76.7] \\
    Article recall, subject issuers & 65 reference pairs & 92.3\% [86.3,98.3] \\
    Focus precision, every link & 60 of 188 predicted pairs & 31.9\%
    [24.7,39.1] \\
    Focus precision, three mentions & 56 predicted pairs & 66.1\% [54.8,77.3] \\
    Focus recall, three mentions & 65 reference pairs & 56.9\% [44.3,69.5] \\
    Explicit ticker anchors & 300 sampled anchors & 294/300 [95.70\%,99.26\%] \\
    Decorated-name additions & 300 sampled new pairs & 300/300
    [98.74\%,100\%] \\
    \bottomrule
  \end{tabular}
\end{table*}

\begin{table*}[t]
  \centering
  \small
  \caption{Coverage by capture cohort. Each column counts distinct IDs.
    Usable means consistently dated and canonical. Linked and focused
    restrict usable IDs to at least one accepted link and at least one
  focused link, respectively. Staged IDs are the cohort denominators.}
  \label{tab:app-coverage}
  \begin{tabular}{rrrrrr}
    \toprule
    Capture year & Staged & Dated & Usable & Linked & Focused \\
    \midrule
    2016 & 61,324 & 54,579 & 54,422 & 18,085 & 6,381 \\
    2017 & 3,659,035 & 772,273 & 770,001 & 302,957 & 129,627 \\
    2018 & 1,055,267 & 1,030,279 & 1,028,757 & 418,127 & 171,662 \\
    2019 & 1,584,644 & 1,252,003 & 1,250,744 & 537,153 & 229,016 \\
    2020 & 1,578,892 & 1,565,227 & 1,563,051 & 793,549 & 366,484 \\
    2021 & 1,588,551 & 1,577,085 & 1,571,857 & 765,957 & 362,237 \\
    2022 & 1,719,346 & 1,557,848 & 1,557,355 & 762,816 & 400,894 \\
    2023 & 1,420,564 & 1,378,531 & 1,375,864 & 680,572 & 357,587 \\
    2024 & 1,033,012 & 1,016,670 & 1,009,969 & 477,955 & 274,446 \\
    2025 & 852,474 & 834,880 & 830,745 & 364,949 & 197,499 \\
    2026 & 254,178 & 248,101 & 248,074 & 98,170 & 51,764 \\
    \bottomrule
  \end{tabular}
\end{table*}

Of the 2017 captures, 2,886,747 arrived more than seven days after
publication and 15 had publication dates beyond the future tolerance
(Table~\ref{tab:app-coverage}). The UK and Indian Reuters hosts alone
account for 1,297,451 and 1,042,605 of these late captures. The
issuer panel assigns
an article to its publication year and counts it once for each linked
issuer. An issuer-year enters the universe with a dated listing or with
a current listing of unknown start and public float observed in that
year or the previous year. Issuers that meet either condition stay in
the panel even in years
when they have no linked articles.

\begin{table}[t]
  \centering\small
  \caption{Issuer-years clearing 50 articles by publication year and
    annual SEC public-float quintile. Parentheses give within-row
    percentages. Float uses the latest measurement in the current or
    preceding year. Annual rank allocation produces slightly uneven
    pooled quintiles. Of all eligible issuer-years, 9,338 have dated
    listing evidence and 25,443 qualify through a current listing with
  unknown start and observed float.}
  \label{tab:size-year}
  \begin{tabular}{lrrr}\toprule
    Group & Eligible & Usable (\%) & Focused (\%) \\ \midrule
    2017 & 3,240 & 854 (26.4) & 326 (10.1) \\
    2018 & 3,341 & 903 (27.0) & 358 (10.7) \\
    2019 & 3,429 & 1,222 (35.6) & 593 (17.3) \\
    2020 & 3,607 & 1,844 (51.1) & 1,184 (32.8) \\
    2021 & 3,950 & 1,923 (48.7) & 1,209 (30.6) \\
    2022 & 4,088 & 2,048 (50.1) & 1,414 (34.6) \\
    2023 & 4,176 & 2,038 (48.8) & 1,475 (35.3) \\
    2024 & 4,334 & 1,778 (41.0) & 1,228 (28.3) \\
    2025 & 4,616 & 1,308 (28.3) & 854 (18.5) \\
    Lowest & 6,533 & 329 (5.0) & 105 (1.6) \\
    Second & 6,530 & 889 (13.6) & 311 (4.8) \\
    Third & 6,529 & 2,084 (31.9) & 914 (14.0) \\
    Fourth & 6,527 & 3,860 (59.1) & 2,161 (33.1) \\
    Highest & 6,526 & 5,790 (88.7) & 4,561 (69.9) \\
    Unmeasured & 2,136 & 966 (45.2) & 589 (27.6) \\
    \bottomrule
  \end{tabular}
\end{table}

\section{Body embeddings and issuer conditioning}

The production EmbeddingGemma artifact \citep{embeddinggemma2025} stores
768 FP16 coordinates per body with a 2,048-token limit. In the census we
reconciled all IDs, positions and text flags and checked vector norms
and finiteness. All 14,807,287 frozen bodies contained text, so we used
a blank-body fixture to check the zero-vector convention. Whereas the
focus screen counts mentions throughout the captured body, in the
overnight panel the tokenizer truncated 44 of 996 bodies after adding
the document prefix and special tokens. For six of 1,104 article--issuer
rows, the third linked mention ended beyond the text visible to the
encoder.

To evaluate whether an embedding can distinguish two firms in the same
story, the issuer-conditioning pilot requires the model reference to
identify a distinct account for each. A paired success occurs when both
company-conditioned views prefer their own name-masked account. Among
150 held-out stories Luna identified 18 eligible cases, of which 16 fit
every encoder, while Terra found eight eligible stories in the audit.
Since both denominators fell below their planned minima of 60 and 20,
the observed gains from selecting issuer sentences warrant confirmation
on a larger sample (Table~\ref{tab:app-vectors}).

\begin{table*}[t]
  \centering
  \small
  \caption{Paired issuer-account retrieval. Each cell counts stories in
    which both company views prefer their own masked account, out of the
  same 16 eligible and common-fit stories.}
  \label{tab:app-vectors}
  \begin{tabular}{lrrr}
    \toprule
    Encoder & Company header & Selected instruction & Issuer sentences \\
    \midrule
    EmbeddingGemma-300M & 2 & 2 & 10 \\
    OpenSearch neural embed v1 & 2 & 2 & 9 \\
    Qwen3-Embedding-0.6B & 2 & 3 & 10 \\
    \bottomrule
  \end{tabular}
\end{table*}

The acquisition retrieval pilot found 38 directional stories among 80
sampled and made 76 comparisons per condition across 19 independent
deal pairs. The primary gain was $-6.6$ percentage points with a paired
95\% interval of $[-18.4,+5.3]$. Terra agreed on 16 of 20 audited stories.
Contextual sentence pooling reused 25 eligible stories and scored
12/25 for Gemma and 11/25 for OpenSearch, each one case below its
extracted-sentence baseline. Both scored 0/24 on controlled role swaps
where the companies share their sentence mask. OpenSearch mention
pooling scored 4/24. The production vectors support the fixed-encoder
exercise while these issuer-specific experiments remain exploratory.

\section{Study details and sensitivities}
\label{sec:app-studies}

\paragraph{Equity market volatility.}
The July 2025 manuscript and archive v1 supply the reference tracker
and specifications \citep{baker2025ms,baker2025archive}. Six archive
reruns recover the monthly fits at displayed precision. To construct
the CCFinance tracker from 196 publisher-month cells,
we divide each publisher's dictionary-hit shares by their dispersion
over the 40 calibration months through December 2022, average across
publishers and rescale to the calibration mean VIX. We then regress
each common outcome on an intercept and its contemporaneous tracker.
For each of 10,000 bootstrap samples, we draw six-month circular
blocks with replacement until we have 49 observations, trimming the
last block. Both trackers and the common outcome use the same months
in each draw. The 2.5th and 97.5th percentiles of the resulting
differences in squared correlations give the paired interval.
Realized-volatility $R^2$ is 0.336 for CCFinance and 0.569 for
the archived tracker, a gap of $-0.233$ $[-0.311,+0.041]$.

Table~\ref{tab:app-emv} drops March and April 2020 first from the two
fits, then also from the calibration of publisher dispersions and the
VIX scale. With all four publishers, the 47-month VIX gap rounds to
$-0.041$ under both calibrations. Omitting Forbes gives the largest
shortfalls among the four publisher omissions in every specification.
Using only the \emph{Wall Street Journal} gives 49-month $R^2$ values
of 0.110 for VIX and 0.120 for realized volatility. The archive's
monthly benchmark supplies the eleven-paper composite, so the
correlation of 0.397 compares the single-publisher tracker with that
composite. Requiring 15 publication dates drops three sparse cells and
yields a full-calendar VIX gap of $-0.142$. Ending calibration in
December 2021 gives a gap of $-0.167$.

\begin{table*}[t]
  \centering
  \small
  \caption{CCFinance minus archive contemporaneous $R^2$. Each 47-month
    fit omits March and April 2020. Calibration uses either the original
    40 months or the 38 remaining after dropping the same two months.
    Each three-publisher tracker retains 147 cells over the full calendar
  and 141 over the restricted calendar.}
  \label{tab:app-emv}
  \begin{tabular}{lrrrrrr}
    \toprule
    & \multicolumn{3}{c}{VIX} & \multicolumn{3}{c}{Realized volatility} \\
    \cmidrule(lr){2-4}\cmidrule(lr){5-7}
    Fit months & 49 & 47 & 47 & 49 & 47 & 47 \\
    Calibration months & 40 & 40 & 38 & 40 & 40 & 38 \\
    \midrule
    All four publishers & $-.132$ & $-.041$ & $-.041$ & $-.233$ &
    $-.051$ & $-.047$ \\
    Omit Benzinga & $-.098$ & $-.067$ & $-.075$ & $-.115$ & $-.010$ & $-.014$ \\
    Omit Forbes & $-.360$ & $-.233$ & $-.240$ & $-.382$ & $-.072$ & $-.072$ \\
    Omit MarketWatch & $-.149$ & $+.004$ & $+.003$ & $-.284$ &
    $-.050$ & $-.045$ \\
    Omit Wall Street Journal & $-.202$ & $-.017$ & $+.001$ & $-.320$
    & $-.036$ & $-.026$ \\
    \bottomrule
  \end{tabular}
\end{table*}

\paragraph{Bank co-mentions.}
The July 2023 CEBRA manuscript defines the reference nodes and windows
\citep{kazinnik2023ms}. The crosswalk maps dated issuers to 33 of its
34 banks, including foreign-parent proxies. Each unordered mapped pair
counts at most once per eligible article. In the matched design,
syndications form connected story components. Fixed quotas retain
682 stories from 31 hosts per week for the comparison and two placebo
changes. From January to April 2020 the matched network adds 56.3
distinct pairs [5, 83] and 307.6 pair observations [178, 421]. From
July to October 2020 it loses 9.8 distinct pairs [$-46$, 28] while
gaining 45.7 observations [$-117$, 217]. From January to April 2021
it loses 21.8 distinct pairs [$-43$, $-1$] while gaining 202.0
observations [70, 341]. Brackets give the central 95\% of changes
across 1,000 matched-story draws.

The extended calendar requires the same roster and publisher quotas over
14 weeks and keeps 158 stories from 14 hosts per week on 32 eligible
nodes. Nasdaq is outside the common publisher support. Truist is
excluded because combining its predecessors would change the network
unit. Although both January--April 2020 draw ranges cross zero, the mean
increase in connected pairs ranks first by signed change among the
seven periods, just 0.1 pair above the next largest, while the increase
in pair weight ranks second. The ranges measure variation across matched
stories within selected weeks, and the six placebo periods supply the
historical comparison (Table~\ref{tab:app-banks}).

\begin{table*}[t]
  \centering
  \small
  \caption{Bank networks on the 14-week common support. Means and 95\%
    ranges summarize matched-story draws with 158 stories and 32 eligible
    banks per week. Pair weight sums article-level co-mentions. The
  ranges summarize within-week sampling.}
  \label{tab:app-banks}
  \begin{tabular}{lll}
    \toprule
    Calendar comparison & Change in distinct pairs & Change in pair weight \\
    \midrule
    Jan.--Apr.\ 2018 & $+11.3$ [$-13$,30] & $+26.9$ [$-36$,89] \\
    Jul.--Oct.\ 2018 & $-5.2$ [$-24$,16] & $-100.1$ [$-161$,$-42$] \\
    Jan.--Apr.\ 2019 & $+4.6$ [$-18$,22] & $+18.7$ [$-77$,120] \\
    Jul.--Oct.\ 2019 & $-2.5$ [$-41$,23] & $+11.1$ [$-85$,99] \\
    Jan.--Apr.\ 2020 & $+19.1$ [$-25$,59] & $+56.2$ [$-40$,155] \\
    Jul.--Oct.\ 2020 & $+19.0$ [$-13$,52] & $+89.8$ [$-14$,199] \\
    Jan.--Apr.\ 2021 & $-49.5$ [$-71$,$-29$] & $-13.8$ [$-99$,67] \\
    \bottomrule
  \end{tabular}
\end{table*}

\paragraph{Overnight portfolios.}
The fixed panel operationalizes a selected design from arXiv v6 and
reference archive v2 \citep{lopez2025v6,lopez2026archive}, using captured
bodies and frozen Yahoo chart prices in place of proprietary headlines
and CRSP. The clock parser recovered 11,196 first-publication markers.
Of those, 2,738 fell outside an overnight interval and 274 were captured
after cutoff. Among the remaining 8,184 articles, 5,392 had no fixed-roster
link. The focus rule rejected 2,510 on-roster links. Of the 995 links
remaining, five failed a within-session similarity screen, one repeated
a recent title and one failed the price-movement headline screen. The
988 survivors passed dated listing, vector and price checks and add
943 unique bodies and 667 firm-sessions. The same 20 of 32 rolling
windows fit, though the recovered panel changes training and evaluation
rows. Allowing LBFGS up to 1,000 iterations eliminated all 16 baseline
drift warnings. At the original inverse regularization strength $C=1$,
nine of 738 chosen security--session signals changed, leaving 39 both-leg
sessions and mean drift of $-8.73$ bp. Stronger regularization
($C=0.1$) changed 221 signals, leaving six both-leg sessions and
mean drift of $-13.67$ bp. Both refits evaluate the original 277 days.

\begin{table*}[t]
  \centering
  \small
  \caption{Mean Gemma body drift in basis points by portfolio support.
    Each panel uses its own evaluated dates. Net means charge 5 bp one
    way on each active leg at entry and exit. The short leg earns the
    negative of its securities' return. Brackets give 95\% intervals
    from 10,000 circular 21-session block samples of the exchange
    calendar between the first and last evaluated dates, conditional
  on saved predictions.}
  \label{tab:overnight-support}
  \begin{tabular}{lrrlrrl}
    \toprule
    & \multicolumn{3}{c}{Original panel} &
    \multicolumn{3}{c}{Recovered panel} \\
    Support & Days & Gross [95\% interval] & Net & Days & Gross [95\%
    interval] & Net \\
    \midrule
    Any active leg & 154 & $-18.26$ [$-52.15$, $7.36$] & $-30.98$
    & 266 & $-9.88$ [$-27.70$, $7.73$] & $-24.02$ \\
    Both legs & 42 & $-18.18$ [$-93.80$, $30.97$] & $-38.18$
    & 110 & $-6.64$ [$-40.12$, $22.53$] & $-26.64$ \\
    Long leg & 118 & $-5.50$ [$-35.83$, $16.66$] & $-15.50$
    & 194 & $0.18$ [$-17.88$, $17.08$] & $-9.82$ \\
    Short leg & 78 & $-27.72$ [$-65.95$, $3.03$] & $-37.72$
    & 182 & $-14.64$ [$-34.61$, $4.70$] & $-24.64$ \\
    \bottomrule
  \end{tabular}
\end{table*}

On days when the short leg traded, it lost 27.72 bp on average in the
original panel and 14.64 bp in the recovered panel
(Table~\ref{tab:overnight-support}). Across all evaluated dates the
5 bp scenario yielded means of $-17.23$ and $-18.79$ bp, respectively.
Restricting both panels to the original 277 dates, the paired net change
was $-1.27$ bp [$-11.19$,$+9.25$].

The representation probes keep the same 996 baseline stories, issuers,
prices and timing restrictions. Gemma title minus Gemma body mean daily
drift is $+5.72$ bp, with a paired 21-session block interval of
$[-10.67,+21.39]$. MiniLM title minus Gemma title is $-0.26$ bp
$[-18.32,+17.47]$. The cost scenarios charge each active leg twice its
entry notional at the assumed one-way rate and exclude borrow fees.
Mean drift includes all 277 evaluated sessions, with no return or cost
on zero-leg sessions. The negative gross means already imply that
break-even would require a rebate in these scenarios
(Table~\ref{tab:app-drift}).

\begin{table*}[t]
  \centering
  \small
  \caption{Gross and scenario-net open-to-close drift across 277 evaluated
    sessions, in basis points of the unit notional assigned to each
    active leg. Cost columns
  give the assumed one-way cost. Zero-leg sessions enter all means.}
  \label{tab:app-drift}
  \begin{tabular}{lrrrrrr}
    \toprule
    Representation & Zero-leg sessions & Gross & 1 bp & 5 bp & 10 bp & 25 bp \\
    \midrule
    Gemma body & 123 & $-10.15$ & $-11.57$ & $-17.23$ & $-24.30$ & $-45.53$ \\
    Gemma title & 124 & $-4.43$ & $-5.76$ & $-11.07$ & $-17.72$ & $-37.64$ \\
    MiniLM title & 125 & $-4.69$ & $-6.03$ & $-11.40$ & $-18.12$ & $-38.26$ \\
    \bottomrule
  \end{tabular}
\end{table*}

\clearpage
\section{Access and reproduction}

Common Crawl makes its captures available and preserves third-party
rights \citep{commoncrawlterms}. Infini-News documents CC BY 4.0 for
derived metadata and an academic research use condition for bulk text
\citep{lazzaroni2026}. Fine News identifies AGPL-3.0-only for its code
and assigns no common license to underlying stories \citep{solovev2026}.
CCFinance uses Apache-2.0 for its project code, and its selector
reads Fine News Parquet outputs through DuckDB. The Baker
and Lopez-Lira--Tang archives are listed under CC BY 4.0
\citep{baker2025archive,lopez2026archive}. Proprietary inputs such as
Factiva, RavenPack and CRSP have separate access requirements.

Reconstructing the exhibits requires access to the frozen inputs as
well as the code. During review the project code and artifact store are
private, and anonymous checks on 22 November 2026 reached the upstream
release but required authentication at the staging snapshot.
We plan a post-review release
of code and derived records with artifact-specific terms. The historical
publisher registry, capture identifiers, copy decisions, mention offsets
and issuer mappings can be documented independently of the article bodies.
Vector redistribution requires a separate terms review.
Frozen labels permit rescoring, while fresh adjudication requires access
to the corresponding models. Codex assisted with manuscript preparation.

\begin{table}[!ht]\centering\small
  \caption{Inputs needed to reconstruct the exhibits.}
  \label{tab:app-access}
  \begin{tabular}{lp{.55\columnwidth}}\toprule
    Exhibit & Required evidence \\ \midrule
    Corpus & Capture IDs, source frames, publisher registry \\
    Coverage & Dated listings, public float, copy and issuer decisions \\
    Extraction & Frozen predictions, offsets and reference judgments \\
    Retrieval & Body vectors and masked issuer accounts \\
    Volatility & Publisher-month shares, archived tracker, market outcomes \\
    Bank pairs & Bank crosswalk and dated story pairs \\
    Overnight drift & Publication and capture clocks, vectors and
    security prices \\
    \bottomrule
  \end{tabular}
\end{table}